\documentstyle[12pt]{article}
\renewcommand{\theequation}{\thesection.\arabic{equation}}
\renewcommand{\baselinestretch}{1.65}

\newfont{\Mb}{msbm10}                    
\newcommand{\R}{\mbox{\Mb\symbol{82}}}

\newcommand{\rA}{\rightarrow}
\newcommand{\lrA}{\leftrightarrow}

\newcommand{\dA}{\downarrow}

\begin{document}

\renewcommand{\baselinestretch}{1.1}

\title{Limits of the energy-momentum tensor
in general relativity}
\author{
F. M. Paiva\thanks{Departamento F\'{\i}sica Te\'orica, Universidade do
Estado do Rio de Janeiro, Rua S\~ao Fran\-cis\-co Xa\-vier 524 (Maracan\~a),
20550-013 Rio de Janeiro - RJ, Brasil, {\sc internet:
fmpaiva@symbcomp.uerj.br}}
,\ \
M. J. Rebou\c{c}as\thanks{Centro Brasileiro de Pesquisas F\'\i sicas,
Rua Dr.\ Xavier Sigaud 150, 22290-180 Rio de Janeiro~--~RJ, Brazil, {\sc
internet: reboucas@cat.cbpf.br}}
,\ \
G. S. Hall\thanks{Department of Mathematical Sciences, University of
Aberdeen, Aberdeen AB9 2TY, Scotland, U. K., {\sc internet:
gsh@maths.abdn.ac.uk}}
\\ \ \ \&  \ \
M. A. H. MacCallum\thanks{School of Mathematical Sciences, Queen Mary
and Westfield College, Mile End Road, London E1 4NS, U. K., {\sc
internet: M.A.H.MacCallum@qmw.ac.uk}}
}
\date{}
\maketitle
\begin{abstract}

A limiting diagram for the Segre classification of the
energy-momentum tensor is obtained
and discussed in connection with a Penrose  specialization
diagram for the Segre
types. A generalization of the coordinate-free
approach to limits of Paiva {\em et al.} to include non-vacuum space-times is
made. Geroch's work on limits of space-times is also extended. The same
argument also justifies part of the procedure for classification of a
given spacetime using Cartan scalars.
\end{abstract}
{\sc pacs} numbers: 04.20.-q \ \ 04.20.Cv \ \ 04.20.Jb

\section{Introduction} \label{SecInt} \setcounter{equation}{0}

The matter content in general relativity theory is described by a
second order symmetric tensor, the energy-momentum tensor. Under
limiting processes, one would like to know which energy-momentum
tensors might arise. A step in this study is the
investigation of the limits of classes of energy-momentum tensors.  A
classification of this tensor is known according to its Segre type.
It seems, therefore, important to investigate the relations
among the Segre types under limiting processes.

In 1969, Geroch \cite{Geroch1969} studied limits
of space-times (see
also \cite{Schmidt1987}). Among other features, he showed that the Penrose
\cite{Penrose1960} specialization diagram for the Petrov classification
(figure~\ref{PetrovEsp}) is in fact a limiting diagram, in the sense
that under limiting processes only space-times with the same Petrov type
or one of its specializations can be reached. Recently, a coordinate-free
technique for studying the limits of vacuum  space-times was developed and 
the limits of some well known vacuum  solutions were investigated
\cite{PaivaReboucasMacCallum1993}. In this approach the Geroch limiting  
diagram for the Petrov classification was extensively used. 
Afterwards  limits of non-vacuum space-times were 
studied \cite{PaivaRomero1993} and the Penrose specialization diagram
for the Segre classification \cite{Penrose1972,PenroseRindler1986v2,%
Hall1985,SanchezPlebanskiPrzanowski1991} was used.

The main aim of this paper is to build a limiting diagram for the
Segre classification. We also compare our diagram with the Penrose 
specialization diagram for the Segre types, obtained through a
dif\/ferent approach in another context. Moreover this work 
extends Geroch's results on limits of space-times, and
it also generalizes the Paiva {\em et al.}
\cite{PaivaReboucasMacCallum1993} coordinate-free
approach to limits so as to include non-vacuum space-times.

We shall use in this work the concept of limit of a space-time
introduced in reference \cite{PaivaReboucasMacCallum1993}, wherein 
by a limit of a space-time, broadly speaking, we mean a limit of a
family of space-times as some free parameters are taken to a limit. 
For instance, in the one-parameter family of Schwarzschild solutions
each member is a Schwarzschild space-time with a specific value for the
mass parameter $m$. By space-time we understand a real 4-dimensional
dif\/ferential manifold with a metric of signature  $(+ - - -)$ 
together with the attendant structures usually required in general 
relativity theory \cite{HawkingEllis1973}.

We also note that although our main aim is to consider families of
space-times, the same arguments about possible limits apply to the
Segre type along curves within a given spacetime, and thus enable us
to show that the use of these types in the classification by Cartan
scalars satisfies the local constancy requirement stated by
Ehlers\cite{Ehlers}. Correspondingly, when we speak of the Segre type
of a spacetime we mean the Segre type at a generic point ---
the limits of behaviours at special
points must also be limits obtainable from the generic type.

Our major aim in the next section is to present a brief 
summary of Geroch's hereditary properties and discuss some
basic properties of Segre classification required for section
\ref{Hereditary}. In section \ref{Hereditary} we study new 
hereditary properties and build a limiting diagram for the 
Segre classification. In section
\ref{Conclusion} we discuss our results and their applications.

\section{Prerequisites} \label{prereq} \setcounter{equation}{0}

Geroch \cite{Geroch1969} shows that there are some properties that are
inherited by all limits of a family of space-times. These properties
he called hereditary. The first hereditary property devised by Geroch can 
be stated as follows: 

\begin{equation} \label{PropH0}
\begin{minipage}[t]{60ex}
{\bf Hereditary property (Geroch):}\\ 
Let $T$ be a tensor or scalar field built from the
metric and its derivatives. If $T$ is zero for all members of
a family of space-times, it is zero for all limits of this family.
\end{minipage}
\end{equation}

{}From this property we easily conclude that the vanishing of
either the Weyl or Ricci tensor or the curvature scalar
are also hereditary properties. What can be said about the Petrov 
and Segre classifications of those tensors under limiting processes ?

\begin{figure}
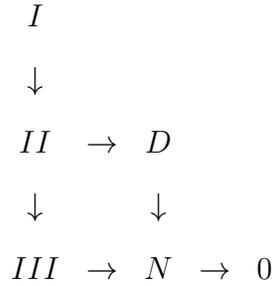
 \[
\begin{array}{ccccc}
I   &     &     &     &   \\
\dA &     &     &     &   \\
II  & \rA & D   &     &   \\
\dA &     & \dA &     &   \\
III & \rA & N   & \rA & 0
\end{array}  \]
\caption{Limiting diagram for the Petrov classification.}
\label{PetrovEsp}
\end{figure}

As far as the Petrov classification is concerned, using the above 
hereditary property, Geroch showed that
although the Petrov type is not a hereditary
property, ``to be at least as specialized as type \ldots'' is. In other
words, the Penrose specialization diagram for the Petrov classification
(figure~\ref{PetrovEsp}) was shown to be a limiting diagram.
For the sake of simplicity, in the limiting diagrams in this paper, we
do not draw arrows between types whenever a compound limit exists.
Thus, in figure~\ref{PetrovEsp}, e.g., the limits $I \rA II \rA D$
imply that the limit $I \rA D$ is allowed.

The Segre classification in general relativity arises from
the eigenvalue problem
$(S^a_b - \lambda \delta^a_b) V^b = 0$
constructed with the trace-free Ricci tensor
$S^a_b \stackrel{def}{=} R^a_b - \frac{1}{4}\delta^a_b R\,.$
By virtue of Einstein's equations, $S^a_b$ and the 
energy-momentum tensor have the same Segre type. 
For an account of the Segre notation and Jordan matrices 
used throughout this article see 
\cite{Hall1984,SantosReboucasTeixeira1995}.
The system above has non-trivial solution only for the 
values of $\lambda$ for which the characteristic polynomial
\begin{equation} %
\mbox{det} \left( S^a_b - \lambda \delta^a_b \right)  \label{PC} %
\end{equation} %
is equal to zero. The fundamental theorem of algebra \cite{Cajori1969} 
ensures that, over the complex field, it can always be factorized as
\begin{equation} \label{PoliCarac}
(\lambda - \lambda_1)^{d_1} \, (\lambda - \lambda_2)^{d_2} \,\,
\cdots \,\, (\lambda - \lambda_r)^{d_r} \,,
\end{equation} 
where $\lambda_i$  ($i=1,2,\, \cdots \, ,r$) 
are the distinct roots of the polynomial (eigenvalues), and 
$d_i$ the corresponding degeneracies. 
To indicate the characteristic polynomial we shall 
introduce a list \{$d_1\, d_2\, \cdots \,d_r$\} of
eigenvalues' degeneracies. However, when complex eigenvalues
exist, instead of a digit to denote the degeneracies we shall
use a letter. So, in table~\ref{SegrePCPM}, for example, 
we have \{$z\bar{z}$11\} and \{2$z\bar{z}$\} instead of
\{1111\} and \{211\}, respectively.

We shall now discuss the minimal polynomial, which will be important in
the derivation of a limiting diagram for the Segre classification. Let
$P$ be a monic matrix polynomial of degree $n$ in $S^a_{\ b}$, i.e., 
\begin{equation}
P = S^n + c_{n-1} \, S^{n-1} + c_{n-2} \, S^{n-2} + 
\, \cdots \, + c_1\, S + c_0\,\delta \,,
\end{equation}
where $\delta$ is the identity matrix and $c_n$ 
are complex numbers.
The polynomial $P$ is said to be the {\em minimal polynomial} 
of $S$ if it is the polynomial of lowest degree in $S$ such 
that $P=0$. It can be shown~\cite{Wilkinson1965}
that the minimal (monic) polynomial is unique and can be 
factorized as
\begin{equation} \label{PoliMin}
(S - \lambda_1 \delta)^{m_1} \,
(S - \lambda_2 \delta)^{m_2} \, \cdots \,
(S - \lambda_r \delta)^{m_r} \,,
\end{equation}
where $m_i$  is the dimension of the       
Jordan submatrix of {\em highest dimension\/} for each eigenvalue
$\lambda_i$. We shall
denote the minimal polynomial in a compact form through a list 
$\| m_1\, m_2\, \cdots\, m_r\,\|$.

We can now find the characteristic and minimal polynomials for
each Segre type. The power of the term corresponding to each eigenvalue in
the characteristic polynomial is the sum of the dimensions of the
Jordan submatrices with the same eigenvalue, whereas in the minimal
polynomial the power is the dimension of the Jordan submatrix of
highest dimension with that eigenvalue. Thus, for example, while
Segre types [(1,11)1], [(21)1] and [31] have the same 
characteristic polynomial, denoted by \{31\}, their corresponding
minimal polynomials are, respectively, given by  $\|11\|$, $\|21\|$
and $\|31\|$. On the other hand, while the Segre types [2(11)] and 
[(21)1] have the same minimal polynomial, denoted by $\|21\|$, the
associated characteristic polynomials are, respectively, given by 
\{22\} and \{31\}. Table~\ref{SegrePCPM}
shows the characteristic and minimal polynomials corresponding to each
Segre type.

\begin{table}
\begin{center} \begin{tabular}{||c||l|l|l|l|l|l|l||} 
\hline\hline 
CP $\rightarrow$  
           &\{1111\}&\{$z\bar{z}$11\}& \{211\}  &\{2$z\bar{z}$\}& \{31\}    & \{22\}      & \{4\}    \\
\cline{1-1}
MP$\ \downarrow$ &&&&&&& \\ \hline\hline 
$\|1111\|$ & [1,111]& [$z\bar z11$]  &          &               &           &             &          \\ \hline 
$\|211\|$  &        &                & [211]    &               &           &             &          \\ \hline 
$\|31\|$   &        &                &          &               & [31]      &             &          \\ \hline 
$\|111\|$  &        &                & [1,1(11)]&[$z\bar z(11)$]&           &             &          \\
           &        &                & [(1,1)11]&               &           &             &          \\ \hline 
$\|21\|$   &        &                &          &               & [(21)1]   & [2(11)]     &          \\ \hline
$\|3\|$    &        &                &          &               &           &             & [(31)]   \\ \hline 
$\|11\|$   &        &                &          &               & [(1,11)1] & [(1,1)(11)] &          \\
           &        &                &          &               & [1,(111)] &             &          \\ \hline 
$\|2\|$    &        &                &          &               &           &             & [(211)]  \\ \hline 
$\|1\|$    &        &                &          &               &           &             & [(1,111)] \\ 
\hline\hline
\end{tabular} \end{center} 
\caption[]{Characteristic and minimal polynomials (columns - CP and rows - MP)
corresponding to the Segre types in general relativity.}
\label{SegrePCPM}
\end{table}

Before closing this section we notice that: (i) the explicit expressions
for the minimal polynomial of a traceless real symmetric tensor defined
on a Lorentzian space can be found in Bona {\em at al.\/} 
\cite{BonaCollMorales1992} and (ii) the minimal polynomial
actually yields a contracted identity for $S^a_{b}$, which allows 
the setting up of a generalized algebraic Rainich condition for this 
tensor \cite{Hall1982}.

\section{Limiting Diagram for Segre Types} \label{Hereditary}
\setcounter{equation}{0}

In this section, we shall first discuss limiting diagrams for both the 
characteristic and minimal polynomials, and combine them
to determine a limiting diagram for the Segre classification.
Afterwards, we shall discuss new hereditary properties to 
refine this first version of the limiting diagram for the
Segre classification.

{}From its definition (\ref{PC}), the characteristic polynomial is built
from the metric and its derivatives. Therefore the eigenvalues, i.e.,
the roots of the characteristic polynomial, are scalars built from 
the metric $g$ and its derivatives. The minimal polynomial, 
in its factorized form (\ref{PoliMin}), is a function only of the trace-free
Ricci tensor and the eigenvalues, thus it is also built from the
metric and its derivatives. Since the characteristic and the minimal
polynomials are, respectively, a scalar and a tensor, the hereditary
property~(\ref{PropH0}) can be applied to both.

The dif\/ference between two roots of the characteristic polynomial is also
a scalar which can built with $g$ and its derivatives. Therefore, at
each degeneracy one scalar becomes zero and thus, by the hereditary property%
~(\ref{PropH0}), under a limiting process, the degeneracy of the
characteristic polynomial either increases or remains the same.
Besides, the real and imaginary parts of complex roots are also scalars 
which can be built with $g$ and its derivatives. Therefore, Segre types with
real roots cannot have as a limit a Segre type
with a non-real root. 
Further, it can be shown (see appendix A) that the limits
\{$z\bar{z}$11\} $\rA$ \{1111\}, \{2$z\bar{z}$\} $\rA$ \{211\} and
\{2$z\bar{z}$\} $\rA$ \{31\} are forbidden.
These results are summarized in the limiting diagram for the
characteristic polynomial shown in figure~\ref{PCPMEsp}. 

Let $P_n$ be the minimal polynomial (of degree $n$) of the trace-free
Ricci tensor $S^a_b$ of a family of space-times. We recall that
by virtue of Einstein's equations $S^a_b$ and the energy-momentum tensor 
$T^a_b$ have the same Segre type. By definition $P_n = 0$
for all members of this family. Thus, according to the hereditary
property~(\ref{PropH0}), $P_n = 0$ for all limits of this family. Since
the minimal polynomial is uniquely defined, for all limits of this family
the minimal polynomial of the trace-free Ricci tensor $S^a_b$ is either 
$P_n$ or a lower degree polynomial. In other words, under limiting processes
of a family of space-times the degree of the minimal polynomial either 
decreases or remains the same.
Besides, from the limiting diagram for the characteristic polynomial in
figure~\ref{PCPMEsp} we notice that the number of distinct eigenvalues either
decreases or remains the same under limiting processes. Taking into
account these two properties, we can work out the limiting diagram for
the minimal polynomial also shown in figure~\ref{PCPMEsp}, where the columns
correspond to the same degree of the minimal polynomial, and the rows
correspond to the same number of eigenvalues. 
 
\begin{figure} \[
\begin{array}{l}
\ \{1111\} \ \ \ \ \ \{z\bar{z}11\}           \\
\ \ \ \ \ \dA \ \ \ \ \swarrow \ \ \ \ \ \dA  \\
\ \{211\} \ \ \ \ \ \ \ \{2z\bar{z}\}         \\
\ \ \ \ \ \dA \ \ \ \ \searrow \ \ \ \ \ \dA  \\
\ \ \{31\} \ \ \ \ \ \ \ \ \{22\}             \\
\ \ \ \ \ \dA \ \ \ \ \swarrow              \\
\ \ \ \{4\} \ \ \ \rA \ \ \ 0  
\\
\\ \multicolumn{1}{c}{\mbox{\tiny CHARACTERISTIC POLYNOMIAL}}
\end{array} \hspace{9ex}
\begin{array}{ccccccc} 
\|1111\| &     &         &     &        &     &        \\
\dA      &     &         &     &        &     &        \\
\|211\|  & \rA & \|111\| &     &        &     &        \\
\dA      &     & \dA     &     &        &     &        \\
\|31\|   & \rA & \|21\|  & \rA & \|11\| &     &        \\
         &     & \dA     &     & \dA    &     &        \\
         &     & \|3\|   & \rA & \|2\|  & \rA & \|1\|
\\ &&&&&& 
\\ \multicolumn{5}{c}{\mbox{\tiny MINIMAL POLYNOMIAL}} 
\end{array}   
\]
\caption
{Limiting diagrams for the characteristic  and
minimal polynomials. The minimal polynomials
$\|22\|$ and $\|4\|$ are not shown since they cannot correspond to
Segre types in general relativity \protect\cite{Hall1984,HawkingEllis1973}.}
\label{PCPMEsp} 
\end{figure}

\begin{figure}
\setlength{\unitlength}{3ex}
\begin{center}
\begin{picture}(12,19)

\put(4,18){\makebox(0,1)[t]{[$z\bar{z}$11]}}
\put(7,18){\makebox(0,1)[t]{[1111]}}
 
\put(7,15){\makebox(0,1)[t]{[211]}}
 
\put(1,12){\makebox(0,1)[t]{[$z\bar{z}$(11)]}}
\put(4.3,12){\makebox(0,1)[t]{[(11)11]}}
\put(10,12){\makebox(0,1)[t]{[31]}}
 
\put(4,9){\makebox(0,1)[t]{[2(11)]}}
\put(10,9){\makebox(0,1)[t]{[(21)1]}}
 
\put(4,6){\makebox(0,1)[t]{[(11)(11)]}}
\put(7,6){\makebox(0,1)[t]{[(31)]}}
\put(10,6){\makebox(0,1)[t]{[(111)1]}}
 
\put(7,3){\makebox(0,1)[t]{[(211)]}}
 
\put(7,0){\makebox(0,1)[t]{[(1111)]}}
 
\put(4,18){\vector(-1,-2){2.3}}
\put(4,18){\vector(4,-3){2.4}}
\put(7,18){\vector(0,-1){1.8}}
 
\put(7,15){\vector(-4,-3){2.4}}
\put(7,15){\vector(4,-3){2.3}}
 
\put(1,12){\vector(4,-3){2.4}}
\put(4,12){\vector(0,-1){1.8}}
\put(4,12){\vector(3,-1){5.5}}
\put(10,12){\vector(0,-1){1.8}}
 
\put(4,9){\vector(0,-1){1.8}}
\put(4,9){\vector(4,-3){2.4}}
\put(10,9){\vector(-4,-3){2.4}}
\put(10,9){\vector(0,-1){1.8}}
 
\put(4,6){\vector(4,-3){2.4}}
\put(7,6){\vector(0,-1){1.8}}
\put(10,6){\vector(-4,-3){2.4}}
 
\put(7,3){\vector(0,-1){1.8}}
 
\end{picture}
\end{center}
\caption{First diagram for the limits of Segre types. This diagram
does not take into account the character of the eigenvectors.}
\label{SegreEsp1} 
\end{figure}
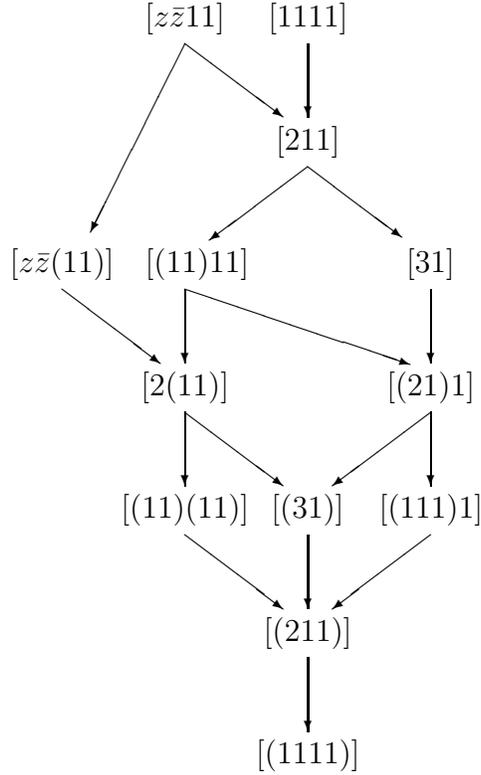

{}From the limiting diagrams for the characteristic and minimal
polynomials in figure~\ref{PCPMEsp}, and table~\ref{SegrePCPM} (which
relates these polynomials to the Segre types) we can draw the
first limiting diagram for the Segre classification, shown in
figure~\ref{SegreEsp1}. Indeed, starting from the limiting
diagram for the minimal polynomial in figure~\ref{PCPMEsp}, we
substitute  for each minimal polynomial the corresponding Segre types
taken from table~\ref{SegrePCPM}. At this point we do not take into
account the character of the eigenvectors. Therefore, we represent the
Segre types [(1,1)11] and [1,1(11)] by [(11)11] and the Segre types
[(1,11)1] and [1,(111)] by [(111)1]. Besides, we notice that in four
situations more than one Segre type is
associated with the same minimal polynomial, namely to the minimal
polynomial $\|21\|$ corresponds
Segre types [(21)1] and [2(11)], to the minimal polynomial $\|11\|$
corresponds Segre types [(111)1] and [(11)(11)], to $\|111\|$
corresponds [11(11)] and [$z\bar{z}$(11)], and finally to the minimal
polynomial $\|1111\|$ corresponds the Segre types 
[1111] and [$z\bar{z}$11]. 

In order to distinguish the Segre type [(21)1] from the type [2(11)], and the
Segre type [(111)1] from [(11)(11)], we shall now consider the
information in the limiting diagram of the characteristic polynomial 
(figure~\ref{PCPMEsp}). We notice that while the Segre types
[(11)11] may have as limit types [2(11)] and [(21)1], the Segre type
[31] may have as limit the type [(21)1] but not the type [2(11)].  Similarly,
the Segre type [2(11)] may have as its limits the types [(31)] and
[(11)(11)], while the Segre type [(21)1] may have the types [(31)] and
[(111)1] as its limits. This completes the limiting diagram
for the Segre classification shown in
figure~\ref{SegreEsp1} (see also ref.\ \cite{Paiva1993}).

It remains to study the consequences of distinguishing 
Segre types which dif\/fer by the character of the eigenvector.
To this end,  we shall now consider the type [(11)11] as 
representing a set of two types, namely [(1,1)11] and [1,1(11)]. 
Furthermore, the type [(111)1] will be looked upon as a set of 
the types [(1,11)1] and [1,(111)]. Firstly we will check whether 
type [(1,1)11] can have [1,1(11)] as its limit and vice-versa, and
similarly whether the type [(1,11)1] can have the type [1,(111)] 
as its limit and reciprocally.
Secondly, we shall find out whether the Segre types which can have as limit 
one of these two Segre set-types can have as limit both members of
the set. Finally, we examine whether the Segre types which can be a limit of
one of these two set-types can be a limit of each member of the
corresponding Segre set-type.

To deal with the questions we have raised in the previous paragraph
we shall introduce three hereditary properties. 
The first one is a corollary of the property~(\ref{PropH0}) and can 
be stated as follows.
\begin{equation} \label{PropHP}
\begin{minipage}[t]{60ex}
{\bf Hereditary property:}\\
Let $E$ be a scalar field built from the metric and its derivatives.
If $E$ is nonzero and $E > 0$ (or $E < 0$) for all members of a 
family of space-times, then $E \geq 0$ (or $E \leq 0$) for
all limits of this family.
\end{minipage}
\end{equation}

The second hereditary property (for a proof see appendix B) we shall 
need can be stated as follows
\begin{equation} \label{PropHNull}
\begin{minipage}[t]{60ex}
{\bf Hereditary property:}\\
If the trace-free Ricci tensor has a null eigenvector for all members of
a family of space-times, then all limits of this family will also have
a null eigenvector.
\end{minipage}
\end{equation}

The Segre type [1,(111)] cannot be obtained as a limit of the type
[(1,11)1] because the latter has null eigenvectors and the former not.
The Segre type [(1,11)1] cannot be obtained as a limit of the type
[1,(111)] because this would involve a change in sign of the invariant 
in the Ludwig-Scanlan classification (types $A_1$ and $A_2$, cf.\
\cite{LudwigScanlan1971,CradeHall1982}). We note that this invariant
also appears in Seixas~\cite{Seixas1991} in a dif\/ferent context.

Again, a consideration of the invariants in the Ludwig-Scanlan 
classification types $B$ and $C$ together with the remark concerning
the associated quartic curve classification of Penrose
~\cite{Penrose1972,CradeHall1982} that a curve of Penrose type $B$
cannot have as its limit a Penrose curve of type $CC$ and vice-versa,
shows that the limits [(1,1)11] $\rA$ [1,1(11)], [1,1(11)] 
$\rA$ [(1,1)11] and [(1,1)11] $\rA$  [2(11)] are forbidden.

In the diagram of figure~\ref{SegreEsp1} the Segre type [211] could
have as a limit the generic set-type [(11)11]. Nevertheless, from the
above property~(\ref{PropHNull}) we verify that [211] can have the 
type [(1,1)11)] as limit, 
but it cannot have as limit the type [1,1(11)]. Similarly, in
figure~\ref{SegreEsp1} the
Segre types [(21)1], [31], [(1,1)11] and [211] could have as limit the
generic set-type [1(111)]. However, the above hereditary property
shows that they can have as limit the type [(1,11)1] but not the type
[1,(111)].

The last hereditary property (see appendix B for a proof) 
we shall need is   
\begin{equation} \label{PropHC}
\begin{minipage}[t]{60ex}
{\bf Hereditary property:}\\
If the trace-free Ricci tensor has a pair of complex conjugate
eigenvalues for all members of a family of space-times, then all
limits of this family will have either a pair of complex conjugate
eigenvalues or at least a null eigenvector. 
\end{minipage}
\end{equation}

In the diagram of figure~\ref{SegreEsp1} the Segre type [$z\bar{z}$11]
could have as limit the generic set-types [(11)11] and [1(111)].
However, according to the hereditary properties~(\ref{PropHNull})
and~(\ref{PropHC}) we conclude that although the type [$z\bar{z}$11] can
still have as its limit the types [(1,1)11] and [(1,11)1]), it cannot have
as limit the types [1,1(11)] and [1,(111)].

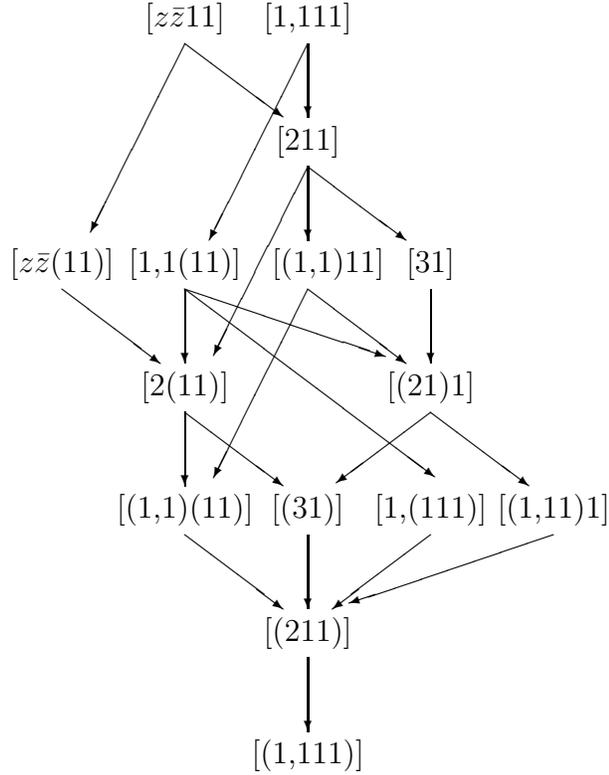
\begin{figure}
\setlength{\unitlength}{3ex}
\begin{center}
\begin{picture}(12,19)

\put(4,18){\makebox(0,1)[t]{[$z\bar{z}$11]}}
\put(7,18){\makebox(0,1)[t]{[1,111]}}

\put(7,15){\makebox(0,1)[t]{[211]}}

\put(1,12){\makebox(0,1)[t]{[$z\bar{z}$(11)]}}
\put(4,12){\makebox(0,1)[t]{[1,1(11)]}}
\put(7.5,12){\makebox(0,1)[t]{[(1,1)11]}}
\put(10,12){\makebox(0,1)[t]{[31]}}

\put(4,9){\makebox(0,1)[t]{[2(11)]}}
\put(10,9){\makebox(0,1)[t]{[(21)1]}}

\put(4,6){\makebox(0,1)[t]{[(1,1)(11)]}}
\put(7,6){\makebox(0,1)[t]{[(31)]}}
\put(10,6){\makebox(0,1)[t]{[1,(111)]}}
\put(13,6){\makebox(0,1)[t]{[(1,11)1]}}

\put(7,3){\makebox(0,1)[t]{[(211)]}}

\put(7,0){\makebox(0,1)[t]{[(1,111)]}}

\put(4,18){\vector(-1,-2){2.3}}
\put(4,18){\vector(4,-3){2.4}}
\put(7,18){\vector(-1,-2){2.4}}
\put(7,18){\vector(0,-1){1.8}}
 
\put(7,15){\vector(-1,-2){2.3}}
\put(7,15){\vector(0,-1){1.8}}
\put(7,15){\vector(4,-3){2.4}}

\put(1,12){\vector(4,-3){2.4}}
\put(4,12){\vector(0,-1){1.8}}
\put(4,12){\vector(4,-3){6.1}}
\put(4,12){\vector(3,-1){4.9}}
\put(7,12){\vector(-1,-2){2.3}}
\put(7,12){\vector(4,-3){2.4}}
\put(10,12){\vector(0,-1){1.8}}

\put(4,9){\vector(0,-1){1.8}}
\put(4,9){\vector(4,-3){2.4}}
\put(10,9){\vector(-4,-3){2.3}}
\put(10,9){\vector(4,-3){2.4}}

\put(4,6){\vector(4,-3){2.4}}
\put(7,6){\vector(0,-1){1.8}}
\put(10,6){\vector(-4,-3){2.4}}
\put(13,6){\vector(-3,-1){5}}

\put(7,3){\vector(0,-1){1.8}}

\end{picture}
\end{center}
\caption{Diagram for the limits of the Segre types of the energy-momentum 
         tensor in general relativity.}
\label{SegreEsp2} 
\end{figure}

Table \ref{Forbidden} summarizes the limits which were forbidden
by the corresponding hereditary properties we have studied in
this section. 
This table together with the diagram in figure~\ref{SegreEsp1} lead 
to the limiting diagram in figure~\ref{SegreEsp2}.

\begin{table}
\begin{center}
\begin{tabular}{|lcl|l|} \hline
{[}1,(111)]      & $\lrA$ & [(1,11)1] \\ 
{[}1,1(11)]      & $\lrA$ & [(1,1)11] \\ 
{[}(1,1)11]      & $\rA$  & [2(11)]   \\ 
{[}211]          & $\rA$  & [1,1(11)] \\ 
{[}(21)1]        & $\rA$  & [1,(111)] \\ 
{[}31]           & $\rA$  & [1,(111)] \\ 
{[}(1,1)11]      & $\rA$  & [1,(111)] \\ 
{[}211]          & $\rA$  & [1,(111)] \\ 
{[}$z\bar{z}$11] & $\rA$  & [1,(111)] \\ 
{[}$z\bar{z}$11] & $\rA$  & [1,1(11)] \\ 
\hline
\end{tabular}
\end{center}
\caption{Limits forbidden by the hereditary properties
\protect\ref{PropHP}, \protect\ref{PropHNull} and \protect\ref{PropHC}.}
\label{Forbidden}
\end{table}

\section{Conclusion} \label{Conclusion} 
\setcounter{equation}{0}

We have built a limiting diagram for the Segre classification 
(figure~\ref{SegreEsp2}) based upon hereditary properties.
To achieve this goal we have extended
Geroch's hereditary properties \cite{Geroch1969}. As a matter of fact, 
we have introduced the properties~(\ref{PropHP}), (\ref{PropHNull}) 
and~(\ref{PropHC}),
worked out limiting diagrams for the characteristic and minimal
polynomials (figure~\ref{PCPMEsp}), and  used them to construct the
limiting diagram shown in figure~\ref{SegreEsp2}.

The {\em limiting\/} diagram for the Segre types we have studied in 
this article essentially coincides with the Penrose {\em specialization\/}
diagram for the Segre classification \cite{Penrose1972,Hall1985}.
As the Penrose specialization is an inverse relation to
deformation \cite{SanchezPlebanskiPrzanowski1991}, the
S\'anchez-Pleba\'nski-Przanowski 
deformation scheme can be obtained from the diagram of figure 4 simply
by reversing the arrows --- limiting and deformation are inverse processes
in a sense. 

Actually, the Penrose diagram is not quite the same as ours since it contains
further subdivisions of the Segre classification. Nevertheless, if one 
rejoins the subdivided cases the two diagrams become identical.

The limiting diagrams of the Petrov and the Segre classification play a
fundamental role in the study of limits of space-times
\cite{PaivaReboucasMacCallum1993,PaivaRomero1993,Paiva1993}, as
briefly discussed in the introduction. The major result of the present 
work, i.e.\  the limiting diagram for the Segre types, also extends
the coordinate-free approach to limits of vacuum space-times  
(where only the Petrov classification is relevant) to 
non-vacuum space-times, where the Segre classification  
plays an essential role. 

The diagram in figure \ref{SegreEsp1}, based on the vanishing of
scalars built with the metric and its derivatives 
together with the further subdivisions corresponding
to the signs of certain scalars, show that the Segre classification
meets Ehlers' requirement\cite{Ehlers} that discrete invariants should
be locally constant.  This amounts to saying that the possible limits
along curves in a given spacetime are the same as those for
one-parameter families of spacetimes. Within a given spacetime, if the
Segre type at a generic point is known, more special types occur only
on submanifolds of lower dimensions, as in the Petrov
classification~\cite{Rendall1988}. 
This justifies one of the steps in the spacetime
classification scheme based on Cartan scalars~\cite{MacCallum1991}.

\section*{Acknowledgment}
FMP and MJR gratefully acknowledge financial assistance from CNPq.
FMP also gratefully acknowledges financial assistance from FAPERJ.

\appendix
\vspace{5mm}
\section*{Appendix A: Forbidden Limits of the  Characteristic Polynomial}
\renewcommand{\theequation}{A.\arabic{equation}}

Our aim in this appendix is to discuss the three forbidden limits
involving the characteristic polynomial with complex roots, which have been 
incorporated in the diagram of the left hand side in figure~\ref{PCPMEsp}.

Let 
\begin{equation}
D = 
(\lambda_1 - \lambda_2)^2
(\lambda_1 - \lambda_3)^2
(\lambda_1 - \lambda_4)^2
(\lambda_2 - \lambda_3)^2
(\lambda_2 - \lambda_4)^2
(\lambda_3 - \lambda_4)^2   \label{escD} 
\end{equation}
be the product of the squares of the dif\/ferences of the 4 roots of the
characteristic polynomial. By direct substitution of 
real, and then complex conjugate roots, one can easily show that 
$D$ is positive for \{1111\} and negative for \{$z\bar{z}$11\}. Since $D$
is built with the metric $g$ and its derivatives, according to the
hereditary property~(\ref{PropHP}) the limit \{$z\bar{z}$11\} $\rA$
\{1111\} is forbidden.

Let now 
\begin{equation}
M = 
(\lambda_1 - \lambda_2)^2
(\lambda_1 - \lambda_3)^2
(\lambda_2 - \lambda_3)^2   \label{escM}
\end{equation}
be the product of the squares of the dif\/ferences of the 3 distinct roots
of the characteristic polynomial \{211\} and \{2$z\bar{z}$\}. Similarly,
by direct substitution of real, and then of complex conjugate roots, one shows
that $M$ is positive for \{211\} and negative for \{2$z\bar{z}$\}.
Again, using the hereditary property~(\ref{PropHP}) the limit 
\{2$z\bar{z}$\} $\rA$ \{211\} is forbidden.

To deal with the limit \{2$z\bar{z}$\} $\rA$ \{31\} consider the 
following  six scalars (built with the metric $g$ and
its derivatives):
$(\lambda_1-\lambda_2)^2$,
$(\lambda_1-\lambda_3)^2$, $(\lambda_1-\lambda_4)^2$,
$(\lambda_2-\lambda_3)^2$, $(\lambda_2-\lambda_4)^2$,
$(\lambda_3-\lambda_4)^2$, 
which are the squares of the dif\/ferences of each pair of roots.
For \{2$z\bar{z}$\} these scalars are such that one is a negative real number,
another is zero, and the remaining four are two equal complex numbers and 
their complex conjugates.
For \{31\} they are three equal real positive numbers,
and three zeros.
Now, from hereditary properties~(\ref{PropH0}) and (\ref{PropHP})
one finds that if the limit \{2$z\bar{z}$\} $\rA$ \{31\} were permitted 
then the scalar which is zero would have to remain zero and the scalar 
which is negative would have to become zero, since \{31\} has no negative 
scalars.
Further, to allow that limit, one out of the 4 complex scalars 
would have to become zero, and the remaining three would have 
to become equal positive real numbers, which clearly is not possible. 
Therefore the limit \{2$z\bar{z}$\} $\rA$ \{31\} is forbidden.

\section*{Appendix B: Proofs of Hereditary Properties}
\renewcommand{\theequation}{B.\arabic{equation}}

Our aim in this appendix is to present proofs of the hereditary 
properties~(\ref{PropHNull}) and~(\ref{PropHC}). 

The property~(\ref{PropHNull}) can be proved as follows.
Let $S$ be the set of real space-time {\em directions\/} at a point $p$ in
the space-time manifold $M$ and let $N$ be the subset of real null
directions at $p$ with respect to the limit metric $g$ at $p$.
Let $R$ be the limit Ricci tensor%
\footnote{Clearly the Ricci tensor has the same Segre type as the 
trace-free Ricci tensor.}
at $p$. If $k$ is a real 
vector at $p$ and $[k]$ the corresponding real direction at $p$,
so that $[k] \in S$, define a map $f: S \mapsto \R$  by
\begin{equation}
f\;: [k]\longmapsto \frac{P_{ab}\,P_{cd}\,h^{ac}\,h^{bd}}%
                         {(k^a\, k^b \,h_{ab})^2} \;,
\end{equation} 
where $h$ is a {\em positive definite\/} metric at $p$ and
\begin{equation}
P_{ab} = k^c \, R_{c\,[a} \,k_{b\,]}\;.
\end{equation}
Then clearly $f([k])$ is well defined on $S$ and vanishes if and
only if $P_{ab}=0$, which is equivalent to $k$ being an eigenvector
of $R$.
Now $S$ ($\approx P^3\, \R$) is compact and $N$, being a closed subset
of $S$, is also compact. So if $R$ has no null eigenvectors at $p$,
$f$ is nowhere zero on $N$. Since $N$ is compact and connected
and $f$ continuous in the natural topologies on $S$ and $\R$ it
follows that $f$ is bounded away from zero. Hence there exists
$\varepsilon > 0$ such that $f > \varepsilon$ on $N$. It follows
that any ``sequence'' of Ricci tensors approaching $R$ (and associated
metrics approaching $g$) must be such that they ``eventually'' have no
real null eigenvectors. The result now follows by contradiction.

To prove the hereditary property~(\ref{PropHC}) it is suf\/ficient to
show that the only types forbidden as a limit by~(\ref{PropHC}) 
(i.e.\ the only remaining Segre types not admitting a null eigenvector)
namely [1,(111)], [1,1(11)] and [1,111] cannot occur as a limit.
Now the fact that [1,111] cannot occur as such a limit can be seen 
either by considering minimal polynomials or by considering the quartic
curves in the Penrose scheme~\cite{Penrose1972}. The other types 
appear in the Ludwig-Scanlan classification scheme as
\begin{eqnarray*}
(a)\quad  [1,(111)] & \qquad \quad A_2\pm      \\
(b)\quad  [1,1(11)] & \qquad \quad B_{3\,f,g}   \\
(c)\quad  [1,1(11)] & \qquad \quad B_{5\,a,b}   \\
(d)\quad  [1,1(11)] & \qquad \quad C_{1}\pm   
\end{eqnarray*}
The possibility (b) can be eliminated by appealing again to the 
Penrose curve classification. To remove the remaining cases (a)
(c) and (d) consider the continuous map $f': S \mapsto \R$  given
in the above notation by
\begin{equation}
f'\;: [k]\longmapsto \frac{R_{ab}\,k^a\,k^b}
                         {h_{ab}\,k^a\, k^b } \;.
\end{equation} 
Then $f'([k])= 0 \Longleftrightarrow R_{ab}\,k^a\,k^b = 0$ and so 
the points of $N$
where $f'$ vanishes are the real null vectors $\ell$ satisfying
$R_{ab}\,\ell^a\,\ell^b =0$ (the ``generalized Debever-Penrose
directions'' \cite{Hall1984}, see also~\cite{CormackHall1981}).
A similar argument to that given regarding the hereditary 
property~(\ref{PropHNull}) together with the facts that types
(a), (c) and (d) above have no such generalized Debever-Penrose
directions whereas the Segre types [$z\bar{z}$11] and 
[$z\bar{z}$(11)] have infinitely many~\cite{Hall1985} completes
the proof.

\vspace{8mm}

\end{document}